\documentclass[preprint,preprintnumbers,amsmath,amssymb]{revtex4}
\usepackage{graphicx}
\usepackage{dcolumn}
\usepackage{bm}
\usepackage{multirow}
\begin{document}
\title{Quantum Adsorption of an Electron to Porous Silicon}
\author{Yanting Zhang}

\author{Dennis P. Clougherty}
\email{dpc@physics.uvm.edu}
\affiliation{
Department of Physics\\
University of Vermont\\
Burlington, VT 05405-0125}

\date{\today}

\begin{abstract}
Using the theory of Zhang and Clougherty [Phys. Rev. Lett. {\bf 108}, 173202 (2012); arXiv:1012.4405],
we provide detailed supporting information concerning the numerical calculations of the probability ${\it s}\left(E\right)$ for a low-energy electron with incident energy $E$ adsorbing to the surface of nanoporous silicon.   
\end{abstract}

\maketitle

\section{introduction}
The numerical values for the rate of sticking $R$ calculated in Zhang and Clougherty (ZC) \cite{yzdc} are obtained by using Eq. (33) in ZC,
\begin{equation}
R=2\pi(\frac{z}{\hbar\omega_{c}})^{\alpha}e^{\alpha}\frac{g^{2}_{1}\rho}{\hbar^{3}} E_{b}\left(\frac{E_b}{E_{b}+\alpha\hbar\omega_{c}}\right)
\label{MFVARate}
\end{equation}
In this supplementary note, we provide numerical values for the variables in this expression for $R$.

The coupling constant $g_{1}$, the strength of bath-assisted particle transitions between the continuum and the bound state, is given by Eq.~(36) in ZC
\begin{equation}
g_{1}=\left\langle k\left|{\partial V_0(x)\over\partial x}\right|b\right\rangle=\int^{\infty}_{0}{\phi^{*}_{k}(x){\partial V_0(x)\over\partial x}\phi_{b}(x)dx}
\label{g1k}
\end{equation}
where $V_{0}(x)$ is the static surface potential with ground state energy denoted by $-E_{b}$. The coupling constant $g_{3}$ is similarly calculated
\begin{equation}
g_{3}=\left\langle b\left|{\partial V_0(x)\over\partial x}\right|b\right\rangle=\int^{\infty}_{0}{\phi^{*}_{b}(x){\partial V_0(x)\over\partial x}\phi_{b}(x)dx}
\label{g3}
\end{equation}

The coupling strength $\alpha$ was introduced in Eq. (5) in ZC
\begin{equation}
J(\omega)\equiv\sum_{q}\frac{g^{2}_{3}}{\hbar^{2}}\sigma^{2}\left(\omega_{q}\right)\delta(\omega-\omega_{q})=\alpha\omega
\label{Jw}
\end{equation}
where the frequency of the excitations is cut off from above by $\omega_{c}$.
The exact form of $\sigma\left(\omega_{q}\right)$ for the case of coupling to Rayleigh phonons is given later in this note (see Eq. (\ref{sigmaQ})). 

Five fundamental variables $g_{1}$, $g_{3}$, $\alpha$, $E_{b}$ and $\omega_{c}$ are needed in order to calculate the numerical sticking rate given by Eq. (\ref{MFVARate}). Once the numerical values of these fundamental variables are obtained, the other variables are calculated with the following formulae, which were also given previously in ZC. 

The constant $\rho$ is calculated by 
\begin{equation}
\rho=\frac{\alpha\hbar^{2}}{g^{2}_{3}}
\end{equation}
The variable $z$ is given by Eq. (28) in ZC
\begin{equation}
z\approx K(\frac{eK}{\hbar\omega_{c}})^{\frac{\alpha}{1-\alpha}}
\label{z}
\end{equation}
where $K$ is given by Eq. (29) in ZC
\begin{equation}
K\approx\frac{(\frac{g_{1}g_{3}\rho}{\hbar}\omega_{c})^{2}}{E+E_{b}+\alpha\hbar\omega_{c}}
\end{equation}

The probability of sticking to the surface is the sticking rate per surface area per unit incoming particle flux,
\begin{equation}
\it s=\sqrt{\frac{2\pi^2m}{E}}R
\end{equation}
In the following sections, we will take the example of an electron sticking to the surface of porous silicon (pSi) (porosity $P=92.9\%$) by emission of a Rayleigh phonon to show how to calculate the five fundamental variables for the specific case of an electron sticking to porous silicon.  The sticking probabilities displayed in Fig.~1 in ZC are calculated using Eq.~(8).

\section{fundamental variables: $g_{1}$, $g_{3}$ \& $E_{b}$} 
We adopt the commonly used model of an attractive image potential with a cut off at $x_{c}=0.05$ \AA, plus a repulsive hard wall \cite{P.A., heinisch}
\begin{equation}
V_{0}(x)=
\begin{cases}
-{q_{e}^2\over 4\pi\epsilon_{0}}{\Lambda_{0}\over x+x_{c}}&\text{ $x>0$} \\
\infty&\text{ $x\le 0$} \\
\end{cases}
\label{Vimage}
\end{equation}
where 
\begin{equation}
\Lambda_{0}=\frac{\kappa-1}{4\left(\kappa+1\right)}
\label{Lambda0}
\end{equation}
and $q_{e}$ is the charge of an electron, $\epsilon_{0}$ is the vacuum permittivity and $\kappa$ is the relative dielectric constant. The average relative dielectric constant for pSi is found from the following three equations \cite{pSi},\\ 
Bruggeman approximation\\
\begin{equation}
P\frac{\kappa_{a}-\kappa_{pSi}}{\kappa_{a}+2\kappa_{pSi}}+(1-P)\frac{\kappa_{cSi}-\kappa_{pSi}}{\kappa_{cSi}+2\kappa_{pSi}}=0
\label{Brugg}
\end{equation} 
Maxwell and Garnett  approximation
\begin{equation}
\frac{\kappa_{pSi}-\kappa_{a}}{\kappa_{pSi}+2\kappa_{a}}=(1-P)\frac{\kappa_{cSi}-\kappa_{a}}{\kappa_{cSi}+2\kappa_{a}}
\label{Maxwell}
\end{equation}
Looyenga approximation
\begin{equation}
\kappa^{\frac{1}{3}}_{pSi}=(1-P)\kappa^{\frac{1}{3}}_{cSi}+P\kappa^{\frac{1}{3}}_{a}
\label{Looy}
\end{equation}
where the relative dielectric constant of bulk crystalline silicon (cSi) and air is respectively $\kappa_{cSi}=11.68$ and $\kappa_{a}=1$. We find for pSi ($P=92.9\%$), $\kappa\approx1.2$ 

There are exact analytical forms for the continuum and bound state wave functions in $V_{0}(x)$. The relevant matrix elements are given by \cite{heinisch}, 
\begin{eqnarray}
g_{1}&=&\left\langle b\left|\frac{dV_{0}(x)}{dx}\right|k\right\rangle=\frac{q_{e}^2}{4\pi\epsilon_{0}}\Lambda_{0}Z^{2}_{b,k}
\label{g1epSi}\\
g_{3}&=&\left\langle b\left|\frac{dV_{0}(x)}{dx}\right|b\right\rangle=\frac{q_{e}^2}{4\pi\epsilon_{0}}\Lambda_{0}Z^{2}_{b,b}
\label{g3epSi}
\end{eqnarray}
where
\begin{eqnarray}
Z^{2}_{b,k}&\equiv&\left\langle b\left|\frac{1}{\left(x+x_{c}\right)^{2}}\right|k\right\rangle\nonumber\\
&=&\frac{\sqrt{k_{0}}}{a^{\frac{3}{2}}_{B}\sqrt{L}N_{\kappa_{b}}}\int^{\infty}_{x^{'}_{c}}dx^{'}W_{\kappa_{b},\frac{1}{2}}(\frac{2\Lambda_{0}x^{'}}{\kappa_{b}})\frac{1}{x^{'2}}\sqrt{\frac{\pi}{1+\tilde{c}^{2}}}\sqrt{2\Lambda_{0}x^{'}}\left[J_{1}\left(2\sqrt{2\Lambda_{0}x^{'}}\right)-\tilde{c}N_{1}\left(2\sqrt{2\Lambda_{0}x^{'}}\right)\right]\nonumber\\
\label{Z2bk}\\
Z^{2}_{b,b}&\equiv&\left\langle b\left|\frac{1}{\left(x+x_{c}\right)^{2}}\right|b\right\rangle\nonumber\\
&=&\frac{1}{a^{2}_{B}N^{2}_{\kappa_{b}}}\int^{\infty}_{x^{'}_{c}}W_{\kappa_{b},\frac{1}{2}}(\frac{2\Lambda_{0}x^{'}}{\kappa_{b}})\frac{1}{x^{'2}}W_{\kappa_{b},\frac{1}{2}}(\frac{2\Lambda_{0}x^{'}}{\kappa_{b}})dx^{'}
\label{Z2bb}
\end{eqnarray} 
Here, $a_{B}$ is Bohr radius and $x^{'}_{c}=x_{c}/a_{B}$ is the lower limit of the integration.  $\tilde{c}$ is given by 
\begin{equation}
\tilde{c}=-\frac{J_{1}\left(2\sqrt{2\Lambda_{0}x^{'}_{c}}\right)}{N_{1}\left(2\sqrt{2\Lambda_{0}x^{'}_{c}}\right)}
\label{tildec}
\end{equation} 
where $J_{1}(x)$ and $N_{1}(x)$ are the Bessel and Neumann functions of order one, respectively. $W_{\kappa_{b},\frac{1}{2}}(\frac{2\Lambda_{0}x}{\kappa_{b}})$ is the Whittaker function, whose expansion in terms of Laguerre polynomials $L_{n}(x)$ for positive non-integer $\kappa_{b}$ is used in the numerical integrations,
\begin{equation}
W_{\kappa_{b},\frac{1}{2}}(x)=\sum_{n=0}\frac{\kappa_{b}(\kappa_{b}-1)e^{-\frac{1}{2}x}L_{n}(x)}{(\kappa_{b}-n)(\kappa_{b}-n-1)\Gamma(2-\kappa_{b})}
\end{equation}
$\kappa_{b}$ is the bound state quantum number and is found from the boundary condition at $x=x_{c}$ on the surface
\begin{equation}
W_{\kappa_{b},\frac{1}{2}}(\frac{2\Lambda_{0}x^{'}_{c}}{\kappa_{b}})=0
\label{b.c.surface}
\end{equation}
The bound state energies are given by  
\begin{equation}
-E_{\kappa_{b}}=-hcR_{y}\frac{\Lambda^{2}_{0}}{\kappa^{2}_{b}}
\label{minusEkappab}
\end{equation}
where $h$ is Planck's constant, $c$ is the speed of light and $R_{y}$ is the Rydberg constant. The ground state energy in $V_{0}(x)$ is $-E_{b}=-7.76$ meV. $k_{0}$ is a dimensionless number calculated from the continuum incident energy $E$,
\begin{equation}
E=hcR_{y}k^{2}_{0}\Lambda^{2}_{0}
\label{equEcontinuum}
\end{equation}
$N_{\kappa_{b}}$ is the reciprocal normalization constant of the bound state wave function. Thus, \begin{equation}
N^{2}_{\kappa_{b}}=\int^{\infty}_{x^{'}_{c}}\left(W_{\kappa_{b},\frac{1}{2}}(\frac{2\Lambda_{0}x^{'}}{\kappa_{b}})\right)^{2}dx^{'}
\label{squaredNkappa}
\end{equation}

We obtain numerical values of the coupling constants $g_{1}$ and $g_{3}$ by numerical integration of Eqs. (\ref{g1epSi}) and (\ref{g3epSi}). We obtain $g_{3}\approx1.3$ meV\AA$^{-1}$ for an electron bound to the surface of pSi ($P=92.9\%$). $g_{1}$ varies with incident energy $E$ and has a range of $57.7\ \mu$eV $\AA^{-1} \leq g_{1} \leq 1.826$ meV$\AA^{-1}$ for incident energies in Fig. 1 in ZC.

\section{fundamental variables: $\omega_{c}$ \& $\alpha$ }
We consider particle-surface coupling through Rayleigh phonons. The cutoff frequency of Rayleigh phonons is approximated by its dispersion relation \cite{flatte}
\begin{equation}
\omega_{c}=\xi c_{t}Q_{c}
\label{omegac}
\end{equation} 
where $\xi$ is the ratio of speed of Rayleigh waves to transverse waves, $c_{t}$ is the transverse speed of sound and $Q_{c}$ is the maximum surface wave vector. We take for pSi ($P=92.9\%$) $\xi\approx0.88$  and $c_{t}/c_{l}\approx0.694$, which follows from cSi data \cite{bouma}. (Ref.\cite{doghmane} shows that $\xi$ and $c_{t}/c_{l}$, where $c_{l}$ is the longitudinal speed of sound, are largely independent of porosity $P$.) $c_{l}$ measured from high pSi with $P\approx80\%$ is approximately $1680$ ms$^{-1}$ \cite{doghmane}. Assuming $c_{l}$ and $c_{t}$ do not vary much for high pSi with $P\gtrsim 80\%$, we obtain $c_{t}\approx 1166$ ms$^{-1}$ for pSi ($P=92.9\%$). We take $Q_{c}=2\pi/a$, where $a=5.43$ \AA\ (the lattice constant of cSi) \cite{kittel}. Hence from equation (\ref{omegac}), we obtain $\omega_{c}\approx 1.19\times 10^{13}$ s$^{-1}$.\\

The exact form for particle-surface phonon coupling is given in Ref. \cite{flatte}. For Rayleigh waves, 
\begin{equation}
\sigma(\omega_{Q})=\left(\frac{\hbar^{2} \tilde{F}^{2}(\sigma)}{S^{2}G\rho_{0}}\right)^{1/4}
\label{sigmaQ}
\end{equation}
where $Q$ is the surface wave vector of Rayleigh waves and $S$ is the surface area of the target. The shear modulus $G$ and the target's density $\rho_{0}$ for pSi ($P=92.9\%$) are approximated by using Ref. \cite{doghmane},
\begin{eqnarray}
\rho_{0}&=&\rho_{cSi}\left(1-P\right)\approx166\ kg\ m^{-3}
\label{rhopSi}\\
G&=&0.482\rho_{0}c^{2}_{l}\approx230 MPa
\label{mupSi}
\end{eqnarray}
We use a mass density of cSi of $\rho_{cSi}=2330$ kg m$^{-3}$ \cite{kittel}.

The constant $\tilde{F}(\sigma)$ is given by \cite{flatte}
\begin{equation}
\tilde{F}^{-1}(\sigma)=8\xi^{-3}\left(\frac{2-\xi^{2}}{2\sqrt{1-\xi^{2}}}-\frac{2-\xi^{2}}{\sqrt{1-(\xi\tau)^{2}}}+\frac{\left(2-\xi^{2}\right)^{2}\left(2-(\xi\tau)^{2}\right)}{8\left(1-(\xi\tau)^{2}\right)^{\frac{3}{2}}}\right)
\label{tildeReciprocalF}
\end{equation}
where $\sigma$ is Poisson's ratio, and $\tau$ is given by
\begin{equation}
\tau=\sqrt{\frac{1-2\sigma}{2-2\sigma}}
\label{tau}
\end{equation}
For Rayleigh waves, $\xi$ is determined by \cite{flatte}
\begin{equation}
\xi^{6}-8\xi^{4}+8\xi^{2}\left(3-2\tau^{2}\right)-16(1-\tau^{2})=0
\label{xi}
\end{equation}
Thus for $\xi=0.88$, we obtain $\sigma\approx0.03$ and $\tilde{F}\left(\sigma\right)\approx0.24$. 

Substituting Eq. (\ref{sigmaQ}) into Eq. (\ref{Jw}) and taking the continuum limit of the sum over all the modes, we obtain for coupling through Rayleigh waves
\begin{equation}
\alpha=\frac{g^{2}_{3}}{\hbar}\frac{1}{2\pi\xi^{2} c^{2}_{t}}\frac{ \tilde{F}(\sigma)}{\sqrt{G\rho_{0}}}
\label{alphaRay}
\end{equation}
where the following vibrational density of states for Rayleigh waves is used
\begin{equation}
D\left(\omega\right)=\frac{S}{2\pi}\frac{\omega}{\xi^{2} c^{2}_{t}}
\label{DOSRay}
\end{equation}
Using the numbers of $g_{3}$, $\xi$, $c_{t}$, $G$, $\rho_{0}$ and $\tilde{F}(\sigma)$ obtained above, we find $\alpha\approx0.008$ for an electron coupled to the surface of pSi through Rayleigh waves.  The numerical values for the sticking probabilities ${\it s}(E)$ contained in Fig.~1 in ZC are now calculated using Eqs.~(1) and (8).

\bibliography{qs-new} 

\begin{thebibliography}{8}
\expandafter\ifx\csname natexlab\endcsname\relax\def\natexlab#1{#1}\fi
\expandafter\ifx\csname bibnamefont\endcsname\relax
  \def\bibnamefont#1{#1}\fi
\expandafter\ifx\csname bibfnamefont\endcsname\relax
  \def\bibfnamefont#1{#1}\fi
\expandafter\ifx\csname citenamefont\endcsname\relax
  \def\citenamefont#1{#1}\fi
\expandafter\ifx\csname url\endcsname\relax
  \def\url#1{\texttt{#1}}\fi
\expandafter\ifx\csname urlprefix\endcsname\relax\def\urlprefix{URL }\fi
\providecommand{\bibinfo}[2]{#2}
\providecommand{\eprint}[2][]{\url{#2}}

\bibitem[{\citenamefont{Zhang and Clougherty}(2012)}]{yzdc}
\bibinfo{author}{\bibfnamefont{Y.}~\bibnamefont{Zhang}} \bibnamefont{and}
  \bibinfo{author}{\bibfnamefont{D.~P.} \bibnamefont{Clougherty}},
  \bibinfo{journal}{Phys.\ Rev.\ Lett.} \textbf{\bibinfo{volume}{108}},
  \bibinfo{pages}{173202} (\bibinfo{year}{2012}).

\bibitem[{\citenamefont{Bruch et~al.}(1997)\citenamefont{Bruch, Cole, and
  Zaremba}}]{P.A.}
\bibinfo{author}{\bibfnamefont{L.~W.} \bibnamefont{Bruch}},
  \bibinfo{author}{\bibfnamefont{M.~W.} \bibnamefont{Cole}}, \bibnamefont{and}
  \bibinfo{author}{\bibfnamefont{E.}~\bibnamefont{Zaremba}},
  \emph{\bibinfo{title}{Physical Adsorption: Forces and Phenomena}}
  (\bibinfo{publisher}{Clarendon Press}, \bibinfo{year}{1997}).

\bibitem[{\citenamefont{Heinisch et~al.}(2010)\citenamefont{Heinisch, Bronold,
  and Fehske}}]{heinisch}
\bibinfo{author}{\bibfnamefont{R.~L.} \bibnamefont{Heinisch}},
  \bibinfo{author}{\bibfnamefont{F.~X.} \bibnamefont{Bronold}},
  \bibnamefont{and} \bibinfo{author}{\bibfnamefont{H.}~\bibnamefont{Fehske}},
  \bibinfo{journal}{Phys.\ Rev.\ B} \textbf{\bibinfo{volume}{81}},
  \bibinfo{pages}{155420} (\bibinfo{year}{2010}).

\bibitem[{\citenamefont{Gaburro et~al.}(2005)\citenamefont{Gaburro, Daldosso,
  and Pavesi}}]{pSi}
\bibinfo{author}{\bibfnamefont{Z.}~\bibnamefont{Gaburro}},
  \bibinfo{author}{\bibfnamefont{N.}~\bibnamefont{Daldosso}}, \bibnamefont{and}
  \bibinfo{author}{\bibfnamefont{L.}~\bibnamefont{Pavesi}},
  \emph{\bibinfo{title}{Porous Silicon}}, Encyclopedia of Condensed Matter
  Physics (\bibinfo{publisher}{Elsevier}, \bibinfo{address}{Amsterdam},
  \bibinfo{year}{2005}).

\bibitem[{\citenamefont{Flatt{\' e} and Kohn}(1991)}]{flatte}
\bibinfo{author}{\bibfnamefont{M.~E.} \bibnamefont{Flatt{\' e}}}
  \bibnamefont{and} \bibinfo{author}{\bibfnamefont{W.}~\bibnamefont{Kohn}},
  \bibinfo{journal}{Phys. Rev. B} \textbf{\bibinfo{volume}{43}},
  \bibinfo{pages}{7422} (\bibinfo{year}{1991}).

\bibitem[{\citenamefont{Boumaiza et~al.}(1999)\citenamefont{Boumaiza, Hadjoub,
  Doghmane, and Deboub}}]{bouma}
\bibinfo{author}{\bibfnamefont{Y.}~\bibnamefont{Boumaiza}},
  \bibinfo{author}{\bibfnamefont{Z.}~\bibnamefont{Hadjoub}},
  \bibinfo{author}{\bibfnamefont{A.}~\bibnamefont{Doghmane}}, \bibnamefont{and}
  \bibinfo{author}{\bibfnamefont{L.}~\bibnamefont{Deboub}},
  \bibinfo{journal}{J. Mater.Sci. Lett.} \textbf{\bibinfo{volume}{18}},
  \bibinfo{pages}{295} (\bibinfo{year}{1999}).

\bibitem[{\citenamefont{Doghmane et~al.}(2006)\citenamefont{Doghmane, Hadjoub,
  Doghmane, and Hadjoub}}]{doghmane}
\bibinfo{author}{\bibfnamefont{A.}~\bibnamefont{Doghmane}},
  \bibinfo{author}{\bibfnamefont{Z.}~\bibnamefont{Hadjoub}},
  \bibinfo{author}{\bibfnamefont{M.}~\bibnamefont{Doghmane}}, \bibnamefont{and}
  \bibinfo{author}{\bibfnamefont{F.}~\bibnamefont{Hadjoub}},
  \bibinfo{journal}{Semiconductor Phys., Quantum Electronics and
  Optoelectronics} \textbf{\bibinfo{volume}{9}}, \bibinfo{pages}{4}
  (\bibinfo{year}{2006}).

\bibitem[{\citenamefont{Kittel}(1996)}]{kittel}
\bibinfo{author}{\bibfnamefont{C.}~\bibnamefont{Kittel}},
  \emph{\bibinfo{title}{Introduction to Solid State Physics}}
  (\bibinfo{publisher}{John Wiley \& Sons, Inc.}, \bibinfo{address}{New York},
  \bibinfo{year}{1996}).

\end{thebibliography}
\end{document}